
\input phyzzx.tex
\baselineskip=14truept
\twelvepoint
\PHYSREV
\vsize 8.5in
\singlespace
\def\ubar{\overline{u}}
\def\cbar{\overline{c}}
\def\tbar{\overline{t}}

\def\bbar{\overline{b}}
\def\qbar{\overline{q}}

\def\bd{B_d^0}
\def\bdb{\overline{B_d^0}}

\def\Dbar{\overline{D}}
\def\to{\rightarrow}

\def\rosner{$^2$}
\def\stone{$^3$}

\def\meko{$^6$}
\def\langacker{$^7$}

\def\growaka{$^{10}$}

\def\donoghue{$^{12}$}

\def\schroder{$^{15}$}

\def\hattori{$^{20}$}
\def\waka{$^{21}$}
\def\alta{$^{22}$}
\def\korner{$^{23}$}
\def\korshu{$^{24}$}
\def\chirality{$^{25}$}

\def\bsw{$^{27}$}
\def\hqet{$^{28}$}
\def\gs{$^{29}$}
\def\rosnerform{$^{30}$}
\def\cleo{$^{31}$}

\def\desh{$^{34}$}

\def\cho{$^{37}$}
\def\rizzo{$^{38}$}

\def\rizz{$^{43}$}

\def\falk{$^{49}$}

\def\lambda{$^{50}$}

\def\grohiok{$^{59}$}

\def\WSref#1&#2&#3(#4){\unskip , #1\bf #2\rm~ (#4) #3}

\def\NPB{{\it Nucl. Phys.~}{\bf B}}
\def\PLB{{\it Phys. Lett.~}{\bf B}}
\def\PRL{\it Phys. Rev. Lett.~}
\def\PRD{{\it Phys. Rev.~}{\bf D}}
\def\ZPC{{\it Z. Phys.~}{\bf C}}
\def\PTP{\it Prog. Theor. Phys.~}
\def\etal{et al.}
\hskip 10truecm
TECHNION-PH-94-6
\hskip 10truecm
February 1994
\vskip 2truecm
\centerline{\bf
THE CHIRALITY OF $b$ QUARK COUPLINGS}
\vskip 2truecm
\centerline{Michael Gronau}
\centerline{\it Department of Physics,
Technion -- Israel Institute of Technology}
\centerline{\it 32000 Haifa, Israel}
\vskip 2truecm
\centerline{\bf Abstract}
\noindent
This review addresses the question of the chirality of
$b$ quark weak couplings from a theoretical and from a purely
phenomenological point of view.
Due to their small magnitude $b$ decay couplings are
subject to possible large corrections from right-handed terms, on the one hand,
and do not affect the
weak interactions of the first two quark families in any observable manner, on
the other hand. This motivated an $SU(2)_L\times SU(2)_R\times
U(1)$ model with purely right-handed $b$-decay
couplings. The model is consistent with experiments, including
recent measurements of a lepton forward-backward asymmetry in
$B\to D^*\ell\nu$ and of
radiative $B$ decays. Parity violating
asymmetries in semileptonic decays of polarized $\Lambda_b$ baryons
produced in $Z^0$ decays
provide a promising way of measuring the chirality of the $b$ to $c$
current.

\vskip 5truecm
\noindent
------------------

\noindent
To appear in {\it  B Decays} (2nd edition), ed. by S. Stone, World
Scientific, Singapore.

\baselineskip=14truept
\vfill\eject
\baselineskip=6truemm
\noindent
{\bf 1. Introduction}
\vskip 4mm

The standard model of electroweak interactions was extremely
successful wherever it has been tested. In the model the charged-current weak
interactions among the three quark families are given by
$$
{\cal L}_W=-{g\over \sqrt{2}}\left(\matrix{\ubar_L&\cbar_L&\tbar_L\cr}\right)
\gamma_{\mu}V\left(\matrix{d_L\cr s_L\cr b_L\cr}\right)W^{\mu}_L~.
\eqno(1)
$$
$q_L={1\over 2}(1-\gamma_5)q$  are left-handed fields. $V$ is the
Cabibbo-Kobayashi-Maskawa\Ref\ckmr{N. Cabibbo\WSref\PRL&10&531(1963);
M. Kobayashi and T. Maskawa \WSref\PTP&49&652(1973).} (CKM) matrix,
the elements of which were determined
in a large variety of experiments.\Ref\rosnerr{J. L. Rosner, this volume.}
In particular, the elements $V_{cb},~V_{ub}$ were measured in certain $B$ meson
decay rates,\refmark{\rosnerr ,~}\Ref\stoner{S. L. Stone, this volume.}
{}~{\it where it was assumed that the relevant quark currents were
left-handed.}
Strictly speaking, the elements of $V$ cannot be determined only from rate
measurements. To determine the chiral structure of the couplings in the first
place one must carry out more detailed experimental studies such as measuring
certain decay distributions. Such V$-$A tests were already made for the
weak
couplings of quarks belonging to the first two families,\Ref\lhr{G. Barbiellini
and C. Santoni, {\it Riv. Nuo. Cim.} {\bf 9(2)} (1986) 1.} and
indirectly for the $V_{tb}$ coupling,\Ref\bisosr{G.
L. Kane and M. Peshkin\WSref\NPB&195&29(1982); D. Schaile and P. M.
Zerwas\WSref\PRD&45&3262(1992).}~before any observation of the
$t$ quark could be made.  On the other hand, until very recently no
chirality test was made for the $b$ decay couplings, $V_{cb}$ and $V_{ub}$,
and their left-handedness remains to be demonstrated. It is this question to
which this chapter is addressed.\Ref\mekor{M. Gronau, {\it Proceedings
of B Factories, The State of the Art in Accelerators, Detectors and Physics},
SLAC, April 1992, ed. D. Hitlin, SLAC report SLAC-400, p.567.}

Because of the small values measured for $V_{cb}$ and $V_{ub}$, right-handed
currents may potentially be relevant in $b$ quark decays more than in any other
process. For instance, present limits on right-handed couplings in nuclear
$\beta$ decays and in $\mu$ decay are at the level of a few percent.
\Ref\langackerr{P. Langacker and S. U. Sankar\WSref\PRD&40&1569(1989), and
references therein.} Since $V_{cb}$ itself is at this level, when compared to
$V_{ud}$ which governs $\beta$ decays, the effects of right-handed currents in
$B$ decays can be highly enhanced. For illustration, imagine that there exists
a
``right-handed" gauge boson ($W_R$) with a mass of 500 GeV, which couples with
full strength via the weak gauge coupling to right-handed quark currents
in $\beta$ decays and in $B$ decays. In this case a V+A term would show up in
$\beta$ decays only at the level of $2-3\%$ in the amplitude, whereas in $B$
decays it would be as large as the V$-$A amplitude or even dominate it.
Thus, $b$
decays are expected to be most sensitive to new right-handed interactions.
Furthermore, due to the smallness of $V_{cb},~V_{ub}$, their effects on other
parameters of the standard model are very small, both via
unitarity of the CKM matrix and via higher order effects. As a consequence $b$
decays are  potentially an ideal place for new right-handed interactions. This
conclusion was reached recently in a model-independent general global analysis
of all allowed new physics effects using constraints  from existing
data.\Ref\londonr{C. P. Burgess, S. Godfrey, H. Konig, D. London and I.
Maksymyk, McGill University report MCGILL-93-12, November 1993.}

There is also an argument, on a purely theoretical level, for large
right-handed $b$ couplings. It is quite
possible that the observed parity violation at low energies
is related at a deeper level
to the hierarchial pattern of the spectrum of quark and lepton masses.
Assuming that both phenomena have a common origin, one may
imagine a situation in which the left-right asymmetry is maximal in the limit
of
vanishing quark and lepton masses, forbidding in this limit right-handed
couplings. In such a scheme right-handed couplings would be highly
suppressed in the
case of the two light quark families and could become large for
heavy
quarks. This idea was demonstrated in a model suggested a few years
ago.\Ref\Fritr{H. Fritzsch,
{\it Les Rencontres de Physique de la Valee d'Aosta, Results and Perspectives
in
Particle Physics}, La Thuile, Italy, March 1991, ed. M. Greco (Editions
Frontieres, 1991), p.177.}

In this paper we will study two complementary aspects of the question of
the chirality of $b$ decay couplings. On the one hand,
in order to bring the problem into focus, we will look at a
theoretical possibility that $b$ couplings are {\it purely} right-handed, an
extreme scenario which cannot yet be excluded. On the other hand, on a
more conservative and phenomenological level, we will investigate
experimental methods to test the
chirality of $b$ decay couplings. The idea that these couplings may be purely
right-handed was demostrated in a model proposed two years ago.
\Ref\growakar{M.
Gronau and S. Wakaizumi\WSref\PRL&68&1814(1992) and in {\it B Decays}, ed. S.
Stone (World Scientific, Singapore, 1992, first edition), p. 479.}~The model
will be described in Section 2, and a few slightly modified versions which
were studied lately will also be mentioned. The  purpose of suggesting this
model
was not so much to offer an alternative scheme to the very successful Standard
Model, but rather to focus on certain soft points of $B$ physics, which must be
strengthened both theoretically and experimentally in order to firmly test the
Standard Model. Indeed, since the model was proposed a few experiments were
made
which have interesting consequences for the model. In Section 3 we will discuss
three recent measurements: The forward-backward lepton asymmetry in $B\to
D^*\ell\nu$, the observation of radiative $B$ decays, and limits on a
right-handed gauge boson from high energy production experiments. These
experimental results will be shown to imply severe constraints on the few
parameters of the model, however they do not rule it out. We will also discuss
the general implications of these experiments on possible
right-handed couplings. In
Section 4 a few methods will be described to measure directly the chirality of
the $b$ to $c$ current. These methods use parity violating asymmetries in
semileptonic decays of polarized $\Lambda_b$'s produced in $Z^0$ decays.
Other methods will be briefly
mentioned. We conclude in Section 5.

\vskip 0.4truecm
\noindent
{\bf 2. A Model with Purely Right-handed $b$ Decays}
\vskip 3mm
\noindent
2.1~{\it The model}
\vskip 3mm

Consider an $SU(2)_L\times SU(2)_R\times U(1)$ model, \Ref\mohar{R. N.
Mohapatra and J. C. Pati\WSref\PRD&11&566(1975);
ibid, {\it Phys. Rev.} {\bf D11} (1975) 2558; R. N. Mohapatra
and G. Senjanovic \WSref\PRD&12&1502(1975).}~in which left-(right-)
handed pairs of up and down-like quarks are doublets of $SU(2)_L$ ($SU(2)_R$),
which couple with $g_L$ ($g_R$) to gauge bosons $W_L$ ($W_R$) and mix through a
mixing matrix $V^L$ ($V^R$). Choose for $V^L$ and $V^R$ the following unitary
forms:\growaka
$$
V^L=\left(\matrix{\cos\theta_c&\sin\theta_c&0\cr
-\sin\theta_c&\cos\theta_c&0\cr
0&0&1\cr}\right)~,\eqno(2)
$$
$$\eqalignno{
V^R &=
\left(\matrix{\exp(-i\phi)&0&0\cr
0&\exp(i\phi)&0\cr
0&0&1\cr}\right)
\left(\matrix{1&0&0\cr
0&{1\over\sqrt{2}}&{1\over\sqrt{2}}\cr
0&-{1\over\sqrt{2}}&{1\over\sqrt{2}}\cr}\right)
\left(\matrix{c&0&s\cr
0&1&0\cr
-s&0&c\cr}\right)
\left(\matrix{c&-s&0\cr
s&c&0\cr
0&0&1\cr}\right)\cr~~~\cr~~~\cr&
=\left(\matrix{c^2\exp(-i\phi)&-cs\exp(-i\phi)&s\exp(-i\phi)\cr
{s(1-c)\over\sqrt{2}}\exp(i\phi)&{(c+s^2)\over\sqrt{2}}\exp(i\phi)
 &{c\over\sqrt{2}}\exp(i\phi)\cr
-{s(1+c)\over\sqrt{2}}&-{(c-s^2)\over\sqrt{2}}
&{c\over\sqrt{2}}\cr}\right)\cr~~~\cr~~~\cr&
\approx \left(\matrix{\exp(-i\phi)&-s\exp(-i\phi)&s\exp(-i\phi)\cr
{s^3\over 2\sqrt{2}}\exp(i\phi)&{1\over\sqrt{2}}\exp(i\phi)
 &{1\over\sqrt{2}}\exp(i\phi)\cr
-s\sqrt{2}&-{1\over\sqrt{2}}
&{1\over\sqrt{2}}\cr}\right)~.&(3)\cr}
$$
$V^L$ mixes only the left-handed quarks of the first two families through the
Cabibbo angle $\theta_c$. $V^R$, on the other hand, mixes the right-handed
quarks of the second and third families maximally (by a $45^0$ angle), and
mixes these fields slightly and equally with the right-handed quark
of the first family. $s$ and $c$ denote the sine and cosine of the
corresponding small mixing angle. The phase $\phi$ is responsible for CP
violation.

In this model, $B$ decays are due to $W_R$ exchange and $W_L-W_R$ mixing. The
dimensionless mass-mixing parameter is denoted by $\zeta$ and we
use the notations\langacker
$$
\zeta_g\equiv{g_R\over g_L}\zeta,~~~\beta_g\equiv{g^2_R\over g^2_L}{M^2_L\over
M^2_R},~~~\xi_g\equiv\sqrt{\beta^2_g+\zeta^2_g}~.
\eqno(4)
$$
$K$ physics implies stringent limits on $W_L-W_R$ mixing,\Ref\donoghuer{J. F.
Donoghue and B. R. Holstein\WSref\PLB&113&382(1982).}
{}~$\vert\zeta_g\vert<0.01$,
although certain theoretical assumptions are invloved in deriving this
limit.\langacker~As we will see in the next section, $B$ physics constrains
this parameter in our model even stronger.  We will therefore neglect it in the
present section. The two parameters $\xi_g,~s$ replace $\vert
V_{cb}\vert,~\vert
V_{ub}\vert$ of the Standard Model in describing the rates of $B$ decays to
charmed and to charmless final states, respectively.

The measured magnitude of $V_{cb}$, obtained from the $B$ decay lifetime and
from its semileptonic decays \rosner$^,$\stone ~is given in our model by
$$
{\xi_g\over\sqrt{2}}=\vert
V_{cb}\vert^{exp}=.038\pm 0.003~.\eqno(5)
$$
Neglecting $\zeta_g$ this implies a rather narrow range for the ``effective"
right-handed gauge boson mass:
$$
325~{\rm GeV}\leq {g_L\over g_R}M_R\leq 375~{\rm GeV}~.\eqno(6)
$$
Note that in general $g_R\not=g_L$, since the discrete L$-$R symmetry of
Left-Right models is broken at
low energies. For $g_R>g_L$ the actual values of $M_R$ are larger than the
numerical values given in (6). We will return to this point when discussing
direct searches for $W_R$ in Section 3. A very strong constraint,
$(g_L/g_R)M_R>1.6~{\rm TeV}$, is known to follow from the measured $K_L-K_S$
mass-difference in Left-Right symmetric models in which
$\vert V^R\vert =\vert V^L\vert$.
\Ref\beallr{G. Beall, M. Bander and A. Soni \WSref\PRL&48&848(1982).} This
limit is avoided in the above model by the very small value of $V^R_{cd}$,
which follows from the equal right-handed mixing of the second and third
families with the first family.

The experimental value of $\vert
V_{ub}/V_{cb}\vert$~\rosner$^,$\stone ~determines the right-handed mixing
angle:
$$
s\sqrt{2}=\vert{V_{ub}\over
V_{cb}}\vert^{exp}=0.08\pm0.03~.\eqno(7)
$$
The phase $\phi$ is determined from the measured value of
$\epsilon_K$, the CP impurity of the neutral $K$ system (see discussion below).
This completely determines $V^R$.

\vfill\eject
\noindent
2.2~{\it Predictions of the model}
\vskip 3mm

Although the structure of weak interactions given in (2)(3) differs
largely from that of the Standard Model, the model is consistent with all
the bulk of information obtained in weak decays of $d,~s,~c$ and $b$ quarks.
Thi
demonstrates that up to now {\it the physics of the two light quark
families has not revealed the deviations from the old Cabibbo scheme
(2) which are predicted in the Standard Model}. That is, the effect of the
third family, induced by the mixing ``angles"$\vert V_{cb}\vert,~\vert
V_{ub}\vert$ in the  unitary CKM matrix, on the physics of the first two
families is too small to have yet been observed. In the Standard Model the $t$
quark has an important contribution in the box-diagram of  $\epsilon_K$, the CP
impurity of the neutral $K$ system.\rosner ~In this Left-Right model
$\epsilon_K
is accounted for by a box diagram with $W_L,~W_R,~c$ and $u$
exchange.\Ref\cpbr{M. Gronau\WSref\PLB&288&90(1992).} The Standard Model
mechanism of $\bd-\bdb$ mixing via the two-$t$ quark box-diagram
\refmark{\rosnerr ,~}\Ref\schroderr{H. Schroeder, this volume.} is not in
opeartion in the Left-Right model. Rather, the mixing is dominated by a
$W_L,~W_R$ box-diagram with $t$ and $c$ exchange: \growaka
$$
x_d\equiv{\Delta M(\bd)\over\Gamma}=0.78({\tau_B\over
1.5ps})({B^{1/2}_B f_B\over 0.2GeV})^2({350GeV\over (g_L/g_R)M_R})^2({s\over
0.06})~.\eqno(8)
$$
This agrees quite nicely with the experimental value.\schroder ~The
corresponding mixing in the $B^0_s$ system is predicted to be enhanced,
$x_s/x_d=(2s\sin\theta_c)^{-1}$, rather similar to the Standard Model.

The model has a
few predictions for $B$ decays, which differ substantially from those of the
Standard Model. The greatest difference occurs in Cabibbo-suppressed decays,
such as $B\to D^{(*)}\Dbar^{(*)}$ (from $b\to c\cbar d$), which are suppressed
in amplitude by $(s/\sqrt{2})^3$ instead of only by $\sin\theta_c$. This
prediction, which assumes that $\zeta_g$ can be neglected, implies that such
decay modes are too rare to ever be observed! Rates at the level expected in
the Standard Model should soon be measured, and would rule out this particular
model. The suppression of processes of the type $B\to D^{(*)}K^{(*)}$ (from
$b\t
c\ubar s$) is less dramatic. Another prediction is that decay amplitudes for
$b\to c\cbar s$, such as $B\to\psi K$, have an extra $1/\sqrt{2}$ suppression
relative to the Standard Model. Such   differences are indistinguishable due to
theoretical uncertainties in calculating nonleptonic decays. The
model also predicts a well-defined pattern of CP asymmetries in $B$ decays, all
of which are given in terms of $\sin\phi$. This pattern can be easily
distinguished from the asymmetries expected in the Standard Model. Predictions
for various final  CP-eigenstates in $B^0_d$ and $B^0_s$ decays are given in
ref.14. For instance, the asymmetries in $\bd\to\psi K_S,~ \bd\to\pi^+\pi^-$
are
given by $\sin\phi,~-\sin\phi$, instead of by $\sin(2\beta),~\sin(2\alpha)$ in
the Standard Model, where $\alpha,~\beta$ are two angles of the CKM unitarity
triangle.\Ref\nirr{Y. Nir and H. R. Quinn, this volume; I. Dunietz, this
volume;
M. Gronau, {\it Proceedings of the Fifth International Symposium on Heavy
Flavor Physics}, Montreal, July 1993, ed. D. B. MacFarlane, to be published.}

One of the perhaps less attractive features of the
model, not so much related to $B$ physics, is the existence of not very heavy
right-handed neutrinos, $m(\nu_R)<m_b-m_c$. These are  required in order to
allo
semileptonic $B$ decays to be mediated by $W_R$ exchange. For massless or very
light $\nu_R$'s ~$\mu$ decay experiments \Ref\muonr{A. E. Jodidio \etal
\WSref\PRD&34&1967(1986); ibid {\bf D37} (1988) 237(E).} already exclude
the $W_R$ mass range of Eq.(6). For massive neutrinos, with $m(\nu_R)>7$ MeV,
th
limits from these experiments become much weaker and are consistent with (6).
Such neutrinos would have to decay sufficiently fast to evade cosmological
energy density constraints. Models with this feature were suggested and
studied in the past.\Ref\cosmor{A. Kumar and R. N.
Mohapatra\WSref\PLB&150&191(1985); W. Buchmueller and D.
Wyler\WSref\PLB&249&458(1990).}

\vskip 3mm
\noindent
2.3~{\it Variants of the model}
\vskip 3mm

A few modifications of this model were suggested recently. In two of these
schemes the $b$ to $c$ coupling is (dominantly) right-handed, whereas $b$ to
$u$
is left-handed as in the Standard Model.  In one of these schemes
\Ref\wylerr{W.
S. Hou and D. Wyler\WSref\PLB&292&364(1992).} the $b$ to $c$ current is
essentially the same as in the above model. The dominantly left-handed $b$ to
$u$ coupling, on the other hand, is given by a  small parameter in $V^L$
($\delta\sim 0.005$)  and  by an even smaller parameter in $V^R$. (In
comparison, note that the two mixing parameters $\sin\theta_c,~s$ in
(2)(3) are not very small). The predictions of this scheme in CP conserving
$B$ decays are quite similar to the above, although $b\to c\cbar d$ is somewhat
less suppressed. In another model,\Ref\hattorir{T. Hattori, T. Hasuike, T.
Hayashi and S. Wakaizumi, University of Tokushima report TOKUSHIMA 93-06,
December 1993.} in which the $b$ to $u$ coupling is left-handed, $V^R$ is
modified to obtain the form    $$ V^R=\left(\matrix{1&0&0\cr 0&S&C\cr
0&C&-S\cr}\right)~,~~~C=0.92~~~(C^2+S^2=1)~.\eqno(9)
$$
The larger value of $V^R_{cb}$ leads to somewhat larger
right-handed gauge boson masses required to account for the measured value of
$V_{cb}$:
$$
370~{\rm GeV}\leq (g_L/g_R)M_R\leq 425~{\rm GeV}~.\eqno(10)
$$
The smaller value
of $V^R_{cs}$ implies a stronger suppression of processes of the type $B\to\psi
K$. $b\to c\cbar d$ is entirely forbidden in this scheme. A few phases are
introduced in $V^R$ to account for $\epsilon_K$ and to lead to certain
predictions for CP asymmetries in $B$ decays.
A less dramatic possibility would be\Ref\wakar{S. Wakaizumi, University of
Tokushima report UT-DP-93-03, October 1993.} that $b$ to $c$ is left-handed
as in the Standard Model,
and that only the very tiny coupling of $b$ to $u$ is
right-handed. Such a scheme does not require a very light $W_R$. Since $M_R$
may be beyond the limits from $\mu$ decay,
the right-handed neutrinos in this scheme do not have to be massive and may
be stable.

\vskip 0.4truecm
\noindent
{\bf 3. Three Recent Measurements related to the $b$ to $c$ Chirality}
\vskip 4mm

The first test of V$-$A in $b$ quark decays, suggested a long time
ago in a pioneering paper by Altarelli {\it et al.},\Ref\altar{G. Altarelli, N.
Cabibbo, G. Corbo, L. Maiani and G. Martinelli,
{\it Nucl. Phys.} {\bf B208} (1982) 365.}~was
to use the inclusive lepton energy spectrum of $B$ decays. The authors studied
the spectrum in a quasi-free decaying quark model, modified by a Gaussian
spectrum of Fermi momentum and by the effects of gluon radiation. They found
that the lepton spectrum becomes slightly softer when the $b$ to $c$ coupling
is  taken to be V+A instead of V$-$A, while assuming V$-$A for the
leptonic current.
For V+A the peak of the energy spectrum is shifted to a lower value by about
200-300 MeV relative to V$-$A. This model calculation depends essentially
on two
parameters: The width of the Gaussian Fermi momenmtum distribution ($p_f$) and
the $c$ to $b$ quark mass ratio ($m_c/m_b$). The rather precise data of
inclusiv
semileptonic decays accumulated so far fits the V$-$A distribution with a
very
high confidence level.\stone ~The difference between this distribution and the
one for a V+A coupling is, however, washed out by the $p_f,~m_c/m_b$
dependence.

Three experiments were carried out recently which are related to the
V$-$A structure of the $b$ to $c$ coupling:

\item{1.} A measurement of the forward-backward asymmetry of the lepton  with
respect to the $D^*$ in $B\to D^*\ell\nu$.

\item{2.} A measurement of $B\to K^*\gamma$ and an upper limit on inclusive
$B\to X_s\gamma$.

\item{3.} Limits from searches of right-handed gauge bosons, $W_R$, from
hadron production experiments.

The implications of these experiments on the chiral structure of the $b$ quark
coupling are sometimes indirect, and by no means unumbiguous. We will study
thei
general significance and their consequences within the left-right model
describe
in the previous section.

\vfill\eject
\noindent
3.1~{\it Forward-backward asymmetry of $\ell$ with respect to the $D^*$ in
$B\to
D^*\ell\nu$}
\vskip 3mm

Koerner and Schuler\Ref\kornerr{J. G. Koerner and G.
A. Schuler\WSref\PLB&226&185(1989).} suggested some time ago to test the $b$ to
$c$ chirality by measuring in $B\to D^*\ell\nu$ the forward-backward asymmetry
of the lepton with respect to the $D^*$ in the $\ell\nu$ center-of-mass frame.
A
simple helicity argument seems to imply that for a V$-$A $b$ to $c$
current,
leptons prefer to move in the hemisphere opposite to the $D^*$, while for V+A
they prefer to move in the same hemisphere. The argument is based on the
idea that in the $B(b\qbar)$ rest-frame the $D^*$ produced by $b$ to $c$ is
dominantly left-handed, since it is made of a preferentially left-handed $c$
quark and a spectator antiquark $\qbar$ which has an equal admixture of both
helicities. To conserve angular momentum in the decay, the virtual $W^-_L$ is
preferentially left-handed in the $B$ rest-frame, and therefore the left-handed
$\ell^-$ prefers to move in a direction opposite to $D^*$ in the
$\ell^-\overline{\nu}$ center-of-mass frame.

While this argument sounds convincing at first sight, it depends on the
assumption that the lepton pair couples to $W_L$. The argument is reversed and
leads to the same directional correlation between $\ell^-$ and $D^*$ when
assuming that in the transition of $b$ to $c$ a virtual $W_R$ is produced,
rather than a $W_L$. In the case of $W_R$ exchange $b$ to $c$ is V+A, the $D^*$
is dominantly right-handed and so is $W^-_R$, and again the right-handed lepton
$\ell^-$ prefers to be emitted in the hemisphere opposite to $D^*$. This simple
argument illustrates the somewhat limited significance of this chirality test,
very similar to the test based on the inclusive lepton spectrum,\alta ~which
als
assumes that the leptonic coupling is V$-$A. In both cases the suggested
test
cannot distinguish the Standard Model from the model of right-handed $b$
couplings described in the previous section.  The two measurements can at most
set limits on $W_L-W_R$ mixing or, in more general terms, on interaction terms
with a V+A $b$ to $c$ coupling and V$-$A leptonic currents.

A calculation of the forward-backward asymmetry in $B\to D^*\ell\nu$ depends on
the dynamical model used to describe the $B$ to $D^*$ transition. Koerner
and Schuller used a specific hadronic model\Ref\korshur{J. G. Koerner and G. A.
Schuler\WSref\ZPC&38&511(1988).} and obtained\korner~ for the Standard Model
an asymmetry of 20$\%$ which is substantial. In order to evaluate the
sensitivit
of the asymmetry to hadron model details and to V+A terms, let us
go over a
more general analysis in some detail.\Ref\chiralityr{M. Gronau and S.
Wakaizumi\WSref\PLB&280&79(1992).}

The four-fold decay distribution of the outgoing $\ell$,$D$ in the cascade
decay  $B\to D^*(\to D\pi)l\nu$ is given in terms of the $B$ to $D^*$ weak
current helicity amplitudes $H_i$ (lepton masses
are neglected):\refmark{\kornerr ,~}\Ref\gilmanr{F. J. Gilman and R. L.
Singleton, Jr.\WSref\PRD&41&142(1990); K. Hagiwara, A. D.
Martin and M. F. Wade\WSref\PLB&228&144(1989).}
$$
{d\Gamma\over dq^2 d\cos\theta_l d\cos\theta^* d\chi}=
$$
$${3G^2_F\over 2048\pi^4}
\vert V_{cb}\vert^2{Kq^2\over M^2_B}B(D^*\to D\pi)
\bigl[(\vert H_+\vert^2+\vert H_-\vert^2)(1+\cos^2\theta_l)\sin^2\theta^*
+4\vert H_0\vert^2\sin^2\theta_l
$$
$$
\cos^2\theta^*
-2{\rm Re}(H_+H^*_-)\sin^2\theta_l\sin^2\theta^*\cos 2\chi-
{\rm Re}((H_++H_-)H^*_0)\sin 2\theta_l\sin 2\theta^*\cos\chi
$$
$$
+2\eta\xi\bigl((\vert H_+\vert^2-\vert H_-\vert^2)
\cos\theta_l\sin^2\theta^*-{\rm Re}((H_+-H_-)H^*_0)\sin\theta_l\sin
2\theta^*\cos\chi\bigr)\bigr]~. \eqno(11)
$$
$q^2$ is the momentum-transfer-squared to the lepton pair; $K$ is the magnitude
of the $D^*$ momentum in the B rest frame; $\theta_l$ is the angle of the
lepton
with respect to the $D^*$ in the $l\nu$ CM frame; $\theta^*$ is the angle of D
relative to B in the $D^*$ rest frame; $\chi$ is the angle between the planes
of
$l\nu$ and $D\pi$ in the B rest frame. $\eta=+1, -1$ describe ~~$l^-\overline
{\nu}, l^+\nu$ final lepton states, respectively. The two cases of $V-A$ and
$V+A$ leptonic currents are given by $\xi=+1, -1$, respectively. $V_{cb}$ is
the $b$ to $c$ coupling, given in the Standard Model by the corresponding CKM
matrix element and in the Left-Right model by the left-hand-side of Eq.(5).
The forward-backward asymmetry is defined by:
$$
a_{FB}={d\Gamma(\theta_l)-d\Gamma(\pi-\theta_l)\over
d\Gamma(\theta_l)+d\Gamma(\pi-\theta_l)}~,
{}~~~~~~~~~~~~~~~~
{\pi\over 2}\leq\theta_l\leq\pi~,
\eqno(12)
$$
where the angles which do not appear in the arguments are integrated over. The
asymmetry projects out the coefficient of the term before last in (11):
$$
a_{FB}=-{3\over 4}\eta\xi{\vert H_+\vert^2-\vert H_-\vert^2\over
\vert H_0\vert^2+\vert H_+\vert^2+\vert H_-\vert^2}~.
\eqno(13)
$$
The numerator and denominator of (13) may be separately integrated over the
remaining variables to obtain the integrated asymmetry:
$$
A_{FB}=-{3\over 4}\eta\xi{\int_0^{q^2_m}(\vert H_+\vert^2-\vert H_-\vert^2)
Kq^2dq^2\over \int_0^{q^2_m}(\vert H_0\vert^2+\vert H_+\vert^2+\vert
H_-\vert^2)
Kq^2 dq^2}~,
$$
$$
q^2_m\equiv (M_B-M_{D^*})^2~,~~~~~~~K\equiv {\sqrt{(M^2_B-M^2_{D^*}-q^2)^2-
4M^2_{D^*}q^2}\over2M_B}~.\eqno(14)
$$

As argued above, one generally expects $\vert H_-\vert>(<)\vert H_+
\vert$ for $V-A ~(V+A)$ $b$ to $c$ coupling, which determines the sign of the
asymmetry. Models of form factors are required in order to calculate the
magnitude of the asymmetry. Such models are usually written for the vector
and axial-vector form factors, which are related to the helicity amplitudes as
follows:
$$\eqalign{
H_{\pm}=(M_B+M_{D^*})A_1(q^2)\mp {2M_BK\over
M_B+M_{D^*}}V(q^2)~,~~~~~~~~~~~~~~~~~~~~~~~~~~~~~~~~~\cr
H_0={1\over
2M_{D^*}\sqrt{q^2}}\bigl[(M^2_B-M^2_{D^*}-q^2)(M_B+M_{D^*})A_1(q^2)
-{4M^2_BK^2\over M_B+M_{D^*}}A_2(q^2)\bigr]~.}\eqno(15)
$$
The hadronic matrix element has the conventional decomposition:
$$
<D^*(k,\epsilon^*)\vert J_{\mu}\vert B(p)>=-{2i\over
M_B+M_{D^*}}V(q^2)\epsilon_{\mu\nu\rho\sigma}\epsilon^{*\nu}p^{\rho}
k^{\sigma}
$$
$$
+(M_B+M_{D^*})A_1(q^2)\epsilon^*_{\mu}
-{1\over M_B+M_{D^*}}A_2(q^2)(\epsilon^*.p)(p+k)_{\mu}+...\eqno(16)
$$
The dots denote a third axial form factor which does not contribute in the
zero-lepton-mass limit.

In hadronic models the helicity amplitudes
of $B\to D^*$ are matched to the helicity structure of the free quark decay,
either at $q^2=0$,\refmark{\korshur ,~}\Ref\bswr{M. Wirbel, B. Stech and M.
Bauer\WSref \ZPC&29&637(1985).} ~or at $q^2=q^2_m$.\Ref\hqetr{N. Isgur and M.
B.
Wise\WSref \PLB&232&113(1989); ibid {\bf B237} (1990) 527; H.
Georgi\WSref\PLB&240&447(1990); A. F. Falk, H.Georgi, B. Grinstein and M.B.
Wise\WSref\NPB&343&1(1990).}$^,$\Ref\gsr{Gilman and Singleton, ref. 26.}
{}~The extrapolation of the form
factors to other values of $q^2$ is given by monopole and quadrupole
expressions,\korshu $^,$\bsw $^,$\gs ~or by a universal velocity transfer
function.\hqet~For the latter case we use a monopole
function.\Ref\rosnerformr{J. L. Rosner\WSref\PRD&42&3732(1990).}
In general one has
$$
V(q^2)={V(0)\over \bigl(1-q^2/M^2_V\bigr)^n}~,~~~~~
A_{1,2}(q^2)={A_{1,2}(0)\over \bigl(1-q^2/M^2_A\bigr)^{n_{1,2}}}~,\eqno(17)
$$
and in the model of ref.28 $A_1$ has besides the pole a multiplicative factor
$\bigl(1-q^2/(M_B+M_{D^*})^2\bigr)$.
The pole parameters in (17) obtain the following values in the four models
when assuming a V$-$A quark current:
$$
\matrix{{\rm
Model}&{\rm Ref.}&V(0)&A_1(0)&A_2(0)&M_V(GeV)&M_A(GeV)&n&n_1&n_2\cr
{\rm BSW}&27 &0.71&0.65&0.69&6.34&6.73&1&1&1&\cr
{\rm KS}&24&0.7&0.7&0.7&6.34&6.34&2&1&2\cr
{\rm GS}&29&0.95&0.69&0.80&6.8&6.8&1&1&1\cr
{\rm HQET}&28,30&0.56&0.56&0.56&4.42&4.42&1&1&1\cr}\eqno(18)
$$
{}From (14)-(18) one finds the following integrated asymmetries in
$\overline{B}^0 (B^-)\to D^*l^-\overline{\nu}$ for V$-$A quark and
lepton currents
($\eta=\xi=1$):\chirality
$$
\matrix{~&{\rm BSW}&{\rm KS}&{\rm GS}&{\rm HQET}\cr
A_{FB}&0.19&0.20&0.24&0.19\cr}\eqno(19)
$$
The effect of a lower cut of 1 GeV/c on the lepton momentum $p_{\ell}$ is to
reduce $A_{FB}$ by about $30\%$ to the range of values 0.13$-$0.17. We conclude
that the asymmetry is rather insensitive to the details of the form factor
model.

The forward-backward asymmetry was measured recently both by CLEO\Ref\cleor{S.
Sanghera \etal (CLEO)\WSref\PRD&47&791(1993).} and by ARGUS.
\Ref\argusr{H. Albrecht \etal (ARGUS)\WSref\ZPC&57&533(1993).}~The two results
obtained with $p_{\ell}>1$ GeV/c,
$$
A_{FB}=\cases {0.14\pm 0.06\pm 0.03~~~&~~~~{\rm CLEO}\cr 0.20\pm 0.08\pm
0.06~~~
&~~~~{\rm ARGUS}~,\cr}\eqno(20)
$$
are in agreement with the above range of calculated asymmetries.

{\it What conclusion should be drawn from this result?}

\noindent
The measured asymmetry is
clearly consistent with the Standard Model prediction and supports the form
factor models. The positive sign, which confirms the simple qualitative
helicity
argument, would change into negative if a V+A $b$ to $c$ coupling were assumed
while the leptons were taken to couple left-handedly ($\xi=+1$). This would
amount, however, to $B$ decays which are due to $W_L-W_R$ mixing alone, which
is
already excluded by existing limits on the mixing. On the other hand, when both
quarks and leptons are assumed to have V+A couplings ($\xi=-1$), as is the case
in the model of right- handed $b$ couplings described in the previous section,
one obtains the same asymmetry as in the Standard Model. In order to evaluate
more generally the significance of the asymmetry measurement, one may consider
a
general $SU(2)_L\times SU(2)_R\times U(1)$ model with arbitrary quark mixing
matrices $V^L$,~ $V^R$ and an arbitrary $W_R$ mass.\chirality ~It is easy to
show then that, {\it as long as one neglects $W_L-W_R$ mixing one always
obtains the Standard Model asymmetry}. This general result follows from having
no interference in decay rates between amplitudes involving V$-$A and
V+A leptonic
currents. Therefore, the asymmetry measurement can only set limits on $W_L-W_R$
mixing. CLEO\cleo ~used their $\theta_{\ell}$ distribution to set an upper
limit
on $W_L-W_R$ mixing, that is on the allowed rates coming from amplitudes in
whic
the quark and leptons couple with opposite chiralities. In the model
of right-handed $b$ couplings this limit implies $(\zeta_g/\beta_g)^2<0.3$,
which, with (5), is seen to be less stringent than previous limits on
$\zeta_g$.\donoghue

\vskip 3mm
\noindent
3.2~{\it Constraint from $b\to s\gamma$ on right-handed $b$ coupling}
\vskip 3mm

Recently CLEO observed exclusive $B\to K^*\gamma$ decays and reported new
limits
on inclusive $B\to X_s\gamma$:\Ref\bsgamar{R. Ammar \etal
(CLEO)\WSref\PRL&71&674(1993).}
$$
BR(B\to K^*\gamma)=(4.5\pm 1.5\pm 0.9)\times 10^{-5},~~
BR(B\to X_s\gamma)<5.4\times 10^{-4}~~(95\%~c.l.)\eqno(21)
$$
The measured exclusive rate may be turned into a conservative lower limit on
the inclusive rate, $BR(B\to X_s\gamma)>0.60\times 10^{-4}$, when taking the
ratio of exclusive-to-inclusive rates to be at most 30$\%$, the largest
estimate
among  various model calculations.\Ref\deshr{N. G. Deshpande, this volume.}
This range of values is consistent with the
Standard Model prediction,\desh ~$BR(B\to X_s\gamma)=(1-4)\times 10^{-4}$.
The agreement between theory and experiment puts limits on
contributions arising from physics beyond the Standard Model, such as in
supersymetric theories and in other two Higgs models.\Ref\Masieror{A. Masiero,
this volume.} We will describe briefly the corresponding limits on right-handed
interactions.

The radiative decay $b\to s\gamma$ was studied
within $SU(2)_L\times SU(2)_R\times U(1)$ models.\Ref\cocor{D.
Cocolicchio, G.
Costa, G. L. Folgi, J. H. Kim and A. Masiero\WSref\PRD&40&1477(1989); G. M.
Asatryan and A. N. Ioannisyan, {\it Sov. J. Nucl. Phys.} {\bf 51} (1990)
858.}$^,$ \Ref\chor{K. Fujikawa and A. Yamada, University of Tokyo report
UT-656, September 1993; P. Cho and M. Misiak,  Caltech report
CALT-68-1893, October 1993;  K. S. Babu, K. Fujikawa and A. Yamada, Bartol
Research Institute report BA-93-69, December 1993.}$^,$\Ref\rizzor{T. G. Rizzo,
SLAC report
SLAC-PUB-6427, January 1994.} ~It was noted that, whereas $W_R$ exchange is not
particularly enhanced relative to $W_L$ exchange, the contribution of terms
which involve $W_L-W_R$ mixing contains, due to its  chiral structure, the
large
enhancement factor $m_t/m_b$. As a consequence, in a Left-Right symmetric model
in which $\vert V^R_{ij}\vert=\vert V^L_{ij}\vert$, this new contribution
become
comparable in magnitude to the Standard Model term for a very small mixing
parameter, $\zeta_g=0.002$.\cho $^,$ \rizzo  ~This value can then be considered
to be an approximate
upper limit, in order not to spoil the agreement between the Standard
Model prediction and experiment. This limit is stronger than the limit
from $K$ decays.\donoghue~On the other hand, since in this model the
right-handed couplings among quarks are equal in magnitude to the corresponding
left-handed couplings and $M^2_R\gg M^2_L$, the contribution of $W_R$ exchange
diagrams is very small, the dependence on $M_R$ is very weak\rizzo ~and no
useful constraint on this mass can be obtained.

{\it Can the $SU(2)_L\times SU(2)_R$ model with purely right-handed $b$
coupling
described in Section 2 account for the $B\to X_s \gamma$ rate, in spite
of the absence of contributions from $W_L$ exchange?}

\noindent
The contribution to the rate from $W_R$ exchange is essentially half of the
Standard Model rate. This follows from the fact that the $B\to
X_s\gamma$ rate which is governed by $(V^R_{tb}V^R_{ts})^2=1/4$ is normalized
by
the inclusive semileptonic decay rate which is proportional to $\vert
V^R_{cb}\vert^2=1/2$. This contribution by itself is larger than the
(theoretical) lower limit, $BR(B\to X_s\gamma)>0.60\times 10^{-4}$, and is
consistent with experiment. In this model the requirement that the enhanced
contribution from terms which involve $W_L-W_R$ mixing is within experimental
bounds sets very stringent limits on $\zeta_g$, $\vert \zeta_g \vert
<0.001$.\rizzo~Observation of inclusive $B\to X_s \gamma$ at a level
near the
present upper limit (21) would require some nonzero contribution from these
terms.

\vskip 3mm
\noindent
3.3~{\it Limits on $M_R$ from searches for right-handed gauge bosons}
\vskip 3mm

The model of right-handed $b$ couplings \growaka ~requires a relatively
light right-handed gauge boson, $325~{\rm GeV}\leq (g_L/g_R)M_R\leq 375~
{\rm GeV}$. This range may be extended up to about $425~{\rm GeV}$ by a slight
variation in the model,\hattori ~and up to about $800~{\rm GeV}$ in a model in
which only $b$ to $u$ is right-handed.\waka ~The value
of $M_R$ depends on $g_R/g_L$. ~$g_R/g_L\not=1$, since the discrete L$-$R
symmetry
is assumed to be broken at low energie. This parameter is generally
expected to be of order one and, most naturally, not to be too different from
one. The simplest GUT extensions prefer values which are somewhat smaller
than one.\Ref\paridar{M. K. Parida and A. Raychaudhuri\WSref\PRD&26&2364(1982);
D. Chang, R. N. Mohapatra and M. K. Parida\WSref\PRD&30&1052(1984).} ~It is
also
possible, however, that the symmetry breaking pattern leads to
$g_R/g_L>1$.\Ref\higgsr{This may be achieved, for instance, by including in the
model discussed in the second paper of Ref.39 pairs of left and right Higgs
triplets ($\Delta_L,~\Delta_R$) which couple to the left-right singlet field
($\eta$) such that they acquire masses at the electroweak scale and at the
scale
of parity restoration, respectively. See Eq.(4.4) of this paper, in which this
would lead to $T_{2R}<T_{2L}$.} A natural range would therefore be
$0.5<g_R/g_L<2$. Methods to measure this ratio were proposed in the
past.\Ref\langacvetr{M. Cvetic, P. Langacker and B.
Kayser\WSref\PRL&68&2871(1992) and references therein.}

Direct searches for $W_R$ bosons in $\overline{p}p$ collisions are
under investigation at
the Tevatron since 1988. The present limit, $M_R>520~{\rm GeV}$,
was obtained by the CDF collaboration for $g_R/g_L=1$.\Ref\cdfr{F. Abe
\etal (CDF)\WSref\PRL&67&2609(1992).} ~This limit was obtained under the
assumptions that $V^R_{ud}=1$, that right-handed neutrinos are not too heavy
($m(\nu_R)<15~{\rm GeV}$), and that the $W_R$ semileptonic decay branching
ratio
is 1/12 per lepton. These assumptions apply to the model presented in
Section 2. While the production cross section is quadratic in $g_R/g_L$,
higher masses $M_R$ are kinematically suppressed. Therefore, the CDF
limit on $M_R$ grows slower than linearly with $g_R/g_L$, reaching the
limit $M_R>630~{\rm GeV}$ at $g_R/g_L=2$.\Ref\rizzr{T. G. Rizzo, Argonne report
ANL-HEP-PR-93-87, November 1993.} ~Comparing this
with the above upper limit from $B$ decays, $(g_L/g_R)M_R\leq 375~{\rm GeV}$,
on
finds that the allowed range is restricted to $(g_R/g_L>1.55,~M_R>580~{\rm
GeV}$
A recent study of the sensitivity of the presently on-going $W_R$ search at the
Tevatron shows that these limits may soon be raised to
the range $g_R/g_L>2,~M_R>750~{\rm GeV}$ if no signal is observed.\rizz~
Such a large
value of $g_R/g_L$ may seem somewhat unnatural, although not entirely
impossible
in models. In the modified version,\hattori ~in which $(g_L/g_R)M_R\leq
425~{\rm
GeV}$, the expected allowed range from this search would be
$g_R/g_L>1.8,~M_R>765~{\rm GeV}$ if no signal is observed. In both models the
expected limits are seen to require quite large values of $g_R/g_L$.

Another place to look for right-handed interactions is in high energy $ep$
scattering at HERA.\Ref\Buchr{W. Buchmueller \etal, {\it Proceedings of the
Physics at HERA Workshop}, eds. W. Buchmueller and G. Ingleman (1991), p.
1001.}
The search reach at HERA, for not very heavy right-handed neutrinos, is
somewhat
lower than the above ranges of the parameters $M_R,~g_R/g_L$.

\vskip 0.4truecm
\noindent
{\bf 4. Measuring the Chirality of the $b$ to $c$ current}

\vskip 3mm
\noindent
4.1~{\it Parity violating asymmetries of type $<\overline{s}.\overline{p}>$
and $<\overline{p}_1.(\overline{p}_2\times\overline{p}_3)>$}
\vskip 3mm

The forward-backward asymmetry in $B\to
D^*\ell\nu$ studied in the previous section is a parity conserving quantity,
of the type $<\overline{p}_l.\overline{p}_{D^*}>$. Therefore, it is
proportional
to the chiralities of both the quark and lepton currents and its sign does not
provide an unambiguous measure of the $b$-coupling chirality. For this purpose
one clearly needs a parity violating observable. Two possibilities of this kind
are measurements of the $\tau$ and stopped $\mu$ polarization in semileptonic
$B
decays by the $\tau$ and $\mu$ decay
distributions.\refmark{\growakar ,~}\Ref\wakr{S. Wakaizumi, Proceedings of the
International Workshop on B Factories,  KEK, November 1992, p. 390.}
{}~This would determine the chirality of the associated lepton current,
thereby distinguishing between $W_L$ and $W_R$ exchange. Such
measurements, which would  search for a $<\overline{s}. \overline{p}>$
correlation, seem unfeasible at present. Stopped muons would require a
dedicated
detector.

Another possibility
would be to search in $B$ decays for a correlation of the type
$<\overline{p}_1.(\overline{p}_2\times\overline{p}_3)>$, which is not only
parity violating but is also odd under time-reversal. To allow for such an
observable, one must have two interfering amplitudes which do not only have
opposite parities but also acquire different final state interaction phases.
The
decay\Ref\hadronbr{T. Browder, K. Honscheid and S. Playfer, this volume.}
{}~$B\to D^* a^+_1(\to\pi^+\pi^+\pi^-)$ seems to be a suitable process to
search
for such a correlation, since the two identical $\pi^+$ can
form with the $\pi^-$ a $\rho^0$ in two different ways.\meko
{}~One would have to
measure the momentum of the $D^*$ perpendicular to the 3$\pi$ decay plane in
the 3$\pi$ center-of-mass frame. To calculate the magnitude of the up-down
asymmetry one would have to assume factorization and to use the $B$ to
$D^*$ form factors measured in $B\to D^*\ell\nu$.\stone ~The vacuum to $a_1$
transition may be taken from the corresponding $\tau$ decay,
$\tau\to a_1\nu$, which was used in a similar way to determine the
chirality of the $\tau\nu$ current.\Ref\taumur{H. Albrecht \etal (ARGUS)
\WSref\PLB&250&164(1990).}~This
calculation would naturally suffer from theoretical uncertainties. However, it
is quite likely that the calculated sign will be independent of these
uncertainties, and will determine directly the chirality of the $b$ to $c$
coupling.

\vfill\eject
\noindent
4.2~{\it Semileptonic decays of polarized $\Lambda_b$'s produced in $e^+e^-\to
Z\to b\bbar$}
\vskip 3mm

The most studied and quite promising method of determining the $b$ quark
chirality relies on the large polarization of $b\bbar$ pairs produced at
the $Z^0$ resonance. The reaction $e^+e^-\to Z\to b\bbar$ is expected to give
rise to $b$ quarks with polarization $P=-0.94$. In the heavy quark symmetry
limit the $\Lambda_b$'s produced in
the subsequent $b$ quark fragmentation retain this polarization.\Ref\polr{F. E.
Close, J. G. Koerner, R. J. N. Phillips and D. J. Summers,  {\it Journ. Phys.}
{\bf G18} (1992) 1716; T. Mannel and G. A. Schuler\WSref\PLB&279&
194(1992).}~Excited $b$-baryons, $\Sigma_b,~\Sigma^*_b$, which decay to
$\Lambda_b$ may
dilute the $\Lambda_b$ polarization by about 30$\%$ if the excited states
are distinct resonances.\Ref\falkr{A.
F. Falk and M. E. Peshkin, SLAC report SLAC-PUB-6311, August 1993,
to be published in Phys.
Rev. D.}
{}~One can then search for spin-momentum correlations between the
$\Lambda_b$ spin direction and the momenta of $\Lambda_b$ decay products. A few
correlations of this kind were calculated recently. We will discuss some of
them
here.


\vskip 3mm
\noindent
4.2.1~{\it Inclusive lepton energy spectrum in semileptonic $\Lambda_b$ decays}
\vskip 3mm

The first suggestion along this line was to measure the inclusive lepton energy
spectrum in semileptonic $\Lambda_b$ decays. Such events were already observed
a
LEP, where they were identified by a cascade $\Lambda$ in association with a
hig
transverse momentum lepton in the same jet.\Ref\lambdar{D. Decamp \etal
(ALEPH)
\WSref\PLB&278&209(1992); P. D. Acton \etal (OPAL)\WSref\PLB&281&394(1992);
ibid {\bf B316} (1993) 435;  P. Abreu \etal (DELPHI)
\WSref\PLB&311&379(1993).} ~The lepton spectrum can be calculated in a free
quark decay model, $b\to c \ell^-\overline{\nu}_{\ell}$, for V$-$A and
V+A quark
and lepton currents:\Ref\inclusiver{J. Amundson, J. L. Rosner, M. Worah and M.
B. Wise, {\it Phys. Rev.} {\bf D47} (1993) 1260.}
$$
{1\over\Gamma}{d^2\Gamma\over dx
d(\cos\psi)}={3x^2(1-\zeta)^2\over f(m^2_c/m^2_b)}\big(1-{2\over 3}x+{2x-1\over
3}\xi P\cos\psi+\zeta[1-{1\over 3}x+{1+x\over 3}\xi P\cos\psi]\big)~,\eqno(22)
$$
where
$$ x\equiv 2E^*_{\ell}/m_b,~~\zeta\equiv m^2_c/[m^2_b(1-x)],~~f(y)\equiv
1-8y+8y^3-12y^2{\rm log}y~.\eqno(23)
$$
$E^*_{\ell},~\psi$ are the lepton energy and its angle with respect to the $b$
quark polarization in the $b$ rest frame.
$\xi=\pm 1$ describe the two cases of V$\pm$A quark and lepton
couplings, corresponding to $W_L$ and $W_R$ exchange. The energy spectrum in
the
forward direction ($\psi=0$) is seen immediately to be harder for V$-$A
than for
V+A, as follows from a simple helicity argument. For comparison, one may also
calculate the spectrum for a V+A $b$ to $c$ coupling and for a
V$-$A lepton currest, corresponding to  $W_L-W_R$ mixing:\Ref\inclr{M.
Gronau and
S. Wakaizumi\WSref\PRD&47&1262(1993).}
$$
{1\over\Gamma}{d^2\Gamma\over dx d(\cos\psi)}=
{6x^2(1-\zeta)^2\over f(m^2_c/m^2_b)}(1-x)(1-P\cos\psi)~.\eqno(24)
$$
This spectrum is harder than in the above two distributions.

The spectra of (22) must be folded by the $b$ to $\Lambda_b$ fragmentation
function which involves large theoretical uncertainties. For simplicity, one
may take a delta function with a laboratory energy of 45 GeV. Using $m_b=5~
{\rm GeV},~m_c=1.66~{\rm GeV},~P=-0.94$ and neglecting the lepton mass,
the resulting
energy distributions were calculated and plotted in Ref.51 for a minimum
transverse lepton momentum of $p^{min}_T=0.8~{\rm GeV}$. A clear
distinction between V$-$A and V+A was noted, where V$-$A gave rise to a
harder
spectrum than V+A.  A high energy lepton tail characterizes left-handed
chirality. The spectrum corresponding to unpolarized $\Lambda_b$'s (which is
common to V$-$A and V+A) lies between these two spectra. In order to
turn this
into a practical chirality test one would have to fold (22) by more
realistic fragmentation functions. One way to control $b$ quark fragmentation
effects and to be able to compare the measured spectrum with the one for
unpolarized $\Lambda_b$'s would be to calibrate the lepton spectrum of
$\Lambda_b$ decays by those of $B$ meson decays, where the $b$ quark is
unpolarized.\Ref\meler{B. Mele and G. Altarelli\WSref\PLB&299&345(1993).}

\vskip 3mm
\noindent
4.2.2~{\it Lepton energy distribution and parity violating asymmetries in
$\Lambda_b\to \Lambda_c\ell\nu$}
\vskip 3mm

Another way to determine the chirality of the $b$ to $c$ coupling is via the
exclusive decay $\Lambda_b\to \Lambda_c\ell\nu$, which can be described in the
framework of the heavy quark effective theory.\hqet ~Events of this type, with
reconstructed $\Lambda_c$'s, were already identified by the ALEPH
collaboration.\Ref\alephr{D. Buskulic \etal (ALEPH)\WSref\PLB&294&145(1992).}
{}~The differential decay rate of polarized $\Lambda_b$'s is conventionally
written in terms of helicity amplitudes: \Ref\lambdacr{F. Hussain, J.G.
Koerner,
and R. Migneron\WSref\PLB&248&406(1990); R. Singleton\WSref\PRD&43&2939(1991);
J. G. Koerner and M. Kraemer\WSref\PLB&275&495(1992); T.
Mannel and G. A. Schuller, ref.48. The subsequent discussion follows ref.59.}
$$
{d^3\Gamma\over
dq^2d\cos\theta_{\ell}d\cos\tilde{\theta}_c}= {G^2_F\vert V_{cb}\vert^2
Kq^2\over 4(4\pi)^3 M^2}f^2 \{
(1-P\cos\tilde{\theta}_c)[{1\over2}(1-\xi\cos\theta_{\ell})^2\vert H_+
\vert^2
$$
$$
+\sin^2\theta_{\ell}\vert h_-\vert^2]+(1+P\cos\tilde{\theta}_c)
[{1\over 2}(1+\xi\cos\theta_{\ell})^2\vert H_-\vert ^2+
\sin^2\theta_{\ell}\vert h_+\vert ^2]\}~.\eqno(25)
$$
\noindent
$f\equiv\sqrt{(E_b+M_b)(E_c+M_c)/4M_bM_c}$, $G_F$ is the Fermi
constant and $q^2$ is the 4-momentum transfer to the
lepton pair. $E_b,~E_c$ ($M_b,~M_c$) are the $\Lambda_b$,~$\Lambda_c$ energies
(masses) in the ($\ell\overline{\nu}$) center-of-mass frame and
$\theta_{\ell}$
is the angle of the lepton in this frame relative to the direction opposite to
the $\Lambda_c$ momentum. $K$ is the momentum of $\Lambda_c$ in the $\Lambda_b$
rest frame and $\tilde{\theta}_c$ is the $\Lambda_c$ angle relative to the
$\Lambda_b$ spin direction in this frame. Again, the two cases of V$-$A
and V+A are
given by $\xi=+1$ and $-1$, respectively.

The helicity amplitudes $H_{\pm}$ and $h_{\pm}$ are related to the invariant
form factors $F_i$ and $G_i$:
$$\eqalign{
H_{\pm}=&-2\sqrt{M_bM_c}(f_-F_1\mp f_0G_1),\cr
h_{\pm}=&-\sqrt{2M_bM_c}\{[f_+F_1+kf_0({F_2\over M_B}+{F_3\over M_c})]
\mp [(2-f_0)G_1-kf_-({G_2\over M_b}+{G_3\over M_c})]\},\cr
f_{\pm}=&{k\over E_c+M_c}\pm {k\over E_b+M_b}~,~~~~~~f_0=1-{k^2\over
(E_b+M_b)(E_c+M_c)}~,~~~~~~k={M_b K\over\sqrt{q^2}}.}
\eqno(26)
$$
$F_i(v.v')$ and $G_i(v.v')$ describe the vector and axial-vector current
matrix elements
$$
\eqalign{
<\Lambda_c(v',s')\vert V_{\mu}\vert\Lambda_b(v,s)>&=\ubar_{\Lambda_c}(v',s')
(F_1\gamma_{\mu}+F_2v_{\mu}+F_3v'_{\mu})u_{\Lambda_b}(v,s)~,\cr
<\Lambda_c(v',s')\vert A_{\mu}\vert\Lambda_b(v,s)>&=\ubar_{\Lambda_c}(v',s')
(G_1\gamma_{\mu}+G_2v_{\mu}+G_3v'_{\mu})\gamma_5 u_{\Lambda_b}(v,s)~.\cr}
\eqno(27)
$$
$v$ and $v'$ ($s$ and $s'$) are the four-velocities (spins) of $\Lambda_b$ and
$\Lambda_c$, respectivley, $v.v'=(M^2_b+M^2_c-q^2)/2M_bM_c$.

In the heavy quark effective theory, with leading $1/m_c$ corrections, the form
factors are given by a single unknown universal function $\eta(v.v')$:
\Ref\hqlambdar{H. Georgi, B. Grinstein and M. B.
Wise\WSref\PLB&252&456(1990); N. Isgur and M. B. Wise\WSref
\NPB&348&276(1991); T.
Mannel, W.
Roberts and Z. Ryzak\WSref\NPB&355&38(1991).}
$$
F_1=\eta(v.v')~,~~~~G_1=\eta(v.v')~,~~~~F_2=G_2=-\eta(v.v')x~,~~~~F_3=G_3=0~.
\eqno(28)
$$
where $x\equiv\overline{\Lambda}/[m_c(1+v.v')],~\overline{\Lambda}\equiv
m_{\Lambda_c}-m_c$,~$m_c$ being the $c$-quark mass. For $\eta(v.v')$ we use a
single-pole structure:\rosnerform
$$
\eta(v.v')={\omega^2_0\over\omega^2_0-2(1-v.v')}~,~~~~~~~\omega_0=0.9~.
\eqno(29)
$$

Eqs.(25)-(29) can be used to calculate various partial distributions. As in
inclusive semileptonic $\Lambda_b$ decays one may calculate the lepton energy
distribution, first in the $\Lambda_b$ rest frame, and then while assuming a
simple fragmentation function, also in the laboratory frame. As in inclusive
decays one finds that the lepton energy distribution is considerably harder
for V$-$A than for V+A.\Ref\hiokir{Z. Hioki\WSref\PLB&303&125(1993);
{\it Z.
Phys.} {\bf C59} (1993) 555; {\it Phys. Rev.} {\bf D48} (1993) 3404.}
A similar distinction
between the two chiralities can be made using the $\Lambda_c$ energy
distribution. The dependence of the distributions on the assumed fragmentation
function would have to be dealt with in ways as described above.

Perhaps a better and
more direct way to test the chirality of the $\cbar b$ current would be to
search for a parity violating asymmetry. A $q^2$-dependent $\Lambda_c$-energy
asymmetry which is manifestly parity odd was recently studied.
\Ref\tanakar{M. Tanaka\WSref\PRD&47&4969(1993).}~It is also possible to
define $q^2$-integrated asymmetries, the sign of which alone would determine
the chirality. For instance, let us define the forward-backward asymmetry of
the
$\Lambda_c$ relative to the $\Lambda_b$ spin in the $\Lambda_b$ rest
frame: \Ref\grohiokr{M. Gronau, T. Hasuike, T. Hattori, Z. Hioki, T. Hayashi
and
S. Wakaizumi, {\it J. Phys.} {\bf G19} (1993) 1987.}
$$
A^c_{FB}\equiv
{\int^1_0{d\Gamma\over d\cos\tilde{\theta}_c}d\cos\tilde{\theta}_c-
\int^0_{-1}{d\Gamma\over
d\cos\tilde{\theta}_c}d\cos\tilde{\theta}_c\over
\int^1_0{d\Gamma\over
d\cos\tilde{\theta}_c}d\cos\tilde{\theta}_c+
\int^0_{-1}{d\Gamma\over
d\cos\tilde{\theta}_c}d\cos\tilde{\theta}_c}~.\eqno(30)
$$
Similarly one may define the forward-backward asymmetry of the lepton $\ell$
relative to the $\Lambda_b$ spin in the $\Lambda_b$ rest frame,
$A^{\ell}_{FB}$.
Using the values $M_b=5.64~{\rm GeV},~M_c=2.28~{\rm GeV},~P=-0.94$ one
obtains for a V$-$A $b$ decay coupling in the heavy quark limit
($\overline{\Lambda}=0$):\grohiok
$$
A^{\ell}_{FB}=0.19~,~~~~~A^c_{FB}=0.17~.
\eqno(31)
$$
These asymmetries may be reduced by the dilution of the
$\Lambda_b$ polarization
through the decay of excited $b$-baryons to $\Lambda_b$.\falk
{}~The two asymmetries change sign when going from a V$-$A to a V+A
$b$-to-$c$ coupling. $\overline{\Lambda}/m_c$ corrections (taking
$\overline{\Lambda}=0.3$) are found to be insignificant, as is also the effect
o
$p^{min}_T=1.0~{\rm GeV}$. In order to measure these asymmetries one would have
to reconstruct the $\Lambda_b$ rest frame. This may be achieved, for instance,
with a suitable vertex detector.

\vskip 3mm
\noindent
4.2.3~{\it Measurement of $\Lambda_c$ polarization through its hadronic decays}
\vskip 3mm

The $\Lambda_c$ in $\Lambda_b\to \Lambda_c\ell\nu$ is expected to be
longitudinally polarized, in a direction which reflects the chirality of the
$b$
to $c$ current. This polarization leads to a spin-momentum correlation with the
momenta of the $\Lambda_c$ hadronic decay products. This idea was applied
very recently
to the
$\Lambda,~\Sigma$ angular distributions in $\Lambda_c\to \Lambda\pi,~\Sigma\pi$
and to the angular distribution of the proton in $\Lambda_c\to pK^-\pi^+$.
\Ref\konigr{B. Koenig, J. G. Koerner and M. Kraemer, DESY report,
DESY 93-135, September 1993, to be published in Phys. Rev. D.}
{}~Using form factors similar to (28)(29) the authors find that the average
longitudinal polarization of the $\Lambda_c$ is about 80$\%$ in the $\Lambda_b$
rest frame and 30-40$\%$ in the laboratory frame. These estimates
assume a maximal $\Lambda_b$ polarization,
disregarding the decay of excited $b$-baryons which decay to
$\Lambda_b$.\falk
{}~One may hope that the
measured branching ratio and decay asymmery parameter in e.g. $\Lambda_c\to
\Lambda\pi$
are sufficiently large for an observation of
the spin-momentum correlation.
\Ref\pvr{P. Avery \etal (CLEO)\WSref\PRL&65&2842(1990); Cornell
Report CLNS-93-1260, November 1993; H. Albrecht \etal
(ARGUS)\WSref\PLB&274&239(1992).}
{}~The sign of the forward-backward asymmetry of the $\Lambda$
with respect to the $\Lambda_c$ direction would by itself distinguish in
this case between V$-$A and V+A $b$ to $c$ couplings.

\vfill\eject
\noindent
{\bf 5. Conclusion}
\vskip 4mm

The V$-$A structure of $b$ decay couplings has not yet been tested. $b$
decays
may be more sensitive than other processes to right-handed interactions due to
the very small couplings through which these decays occur. Unambiguous
tests of V$-$A would have to rely on parity violating observables or on
resulting energy distributions. Polarized
$\Lambda_b$'s produced on the $Z^0$ resonance provide a useful system for
carrying out such measurements in the not too far future. Such measurements
can verify the V$-$A nature of the $b$ to $c$ coupling, set limits on V+A
terms, and
rule out the model presented in Section 2. It is much more difficult
to test the chirality of the tiny $b$ to $u$ coupling.
The particular model of Section 2 can be ruled out also by an
observation of Cabibbo-suppressed decays of the type $B\to D^{(*)}
\Dbar^{(*)}$ at a level expected in the Standard Model.

\vskip 0.4truecm
\noindent
{\bf 6. Acknowledgements}
\vskip 4mm

I am grateful to S. Wakaizumi and his colleagues for collaboration on some of
th
topics discussed here and for extending their warm hospitality at the
University of Tokushima. I wish to thank J. Goldberg, R. N. Mohapatra, T.
Rizzo, J. Rosner and S. Stone for helpful discussions. This work was supported
in part by the U.S.$-$Israel Binational Science Foundation, by the
German$-$Israeli
Foundation for Scientific Research and Development, by the VPR research fund
and
by the Fund for Promotion of Research at the Technion.

\vfill\eject
\refout
\vfill
\eject
\bye